\documentclass[a4paper]{jpconf}
\usepackage{graphicx}
\usepackage{amsmath}
\usepackage{amssymb}
\begin{document}
\title{Astrospheres and Cosmic Rays}

\author{A. Struminsky$^{1,2}$ and A. Sadovski$^{1,3}$}

\address{$^1$ Space Research Institute, 84/32 Profsouznaya str., Moscow, 117997, Russia\\
	 $^2$ Moscow Institute of Physics and Technology, 9 Institutskiy per., Dolgoprudny, Moscow Region, 141701, Russia\\
 $^3$ National Research University Higher School of Economics, 20 Myasnitskaya ulitsa, Moscow, 101000, Russia}

\ead{asadovsk@iki.rssi.ru}

\begin{abstract}
An ionization plays a key role in formation of stars, planets and their 
atmospheres. Cosmic rays (CR) are the main source of the ionization, 
therefore it is important to know and be able to estimate fluxes of galactic 
and stellar cosmic rays (GCR and SCR) at different stages of evolution of 
stars and planetary systems. Radiation conditions close to exoplanets might 
be important for creation and development of life. We present a 
review of the current state of the problem of astrospheres and their 
interactions with GCR and SCR. We pay special attention to 
estimates of radiation conditions near exoplanets recently discovered in a 
habitable zone of their hosting stars.
\end{abstract}

\section{Introduction---astrophysics of low energy CR}
Traditionally the astrophysics of cosmic rays (OG section of CR conferences) 
considers CR above 10~GeV since CR of lower energies have their flux 
modulated in the heliosphere by the solar magnetic field and the solar wind, some part of low CR rays are 
accelerated inside the heliosphere in solar flares or by shock waves driven 
by Coronal Mass Ejections (CME). Problems of low energy CR physics were 
discussed at SH (Sun and Heliosphere) section of CR conferences \cite{Ginzburg-etal}. Recent years are the birth time of low energy CR astrophysics, which 
is tightly connected to astrochemistry, radioastronomy and solar physics. 
At present the astrophysics of low energy CR consider CR in three locations---molecular clouds, 
protoplanetary discs and astrospheres.

The cosmic-rays dominate in the ionization and heating of the interstellar medium (ISM). Recent measurements, made beyond the heliopause by the Voyager I spacecraft, have provided data of lower-energy cosmic-ray protons and electrons down to energies about 3~MeV \cite{Cummings2016}. Nevertheless, it remains 
unclear whether the cosmic-ray fluxes reach their unmodulated values even at 
the current location of Voyager I. However, the ionization cross-sections for atomic and molecular hydrogen have maximum at 0.01~MeV, so an
extrapolation to unobserved energies is still required to determine the 
implied cosmic rays ionisation rate,
which remains quite uncertain even in the solar neighbourhood 
\cite{NeufeldWolfire2017}. 

Remarkably, the spectra of both proton and electron CRs in the local 
interstellar medium (ISM) at least down to particle energy of a few MeVs are 
now known with some confidence, thanks to the recent data collected by the 
Voyager probe at large distances from the Sun \cite{Cummings2016}. Whether 
or not such spectra are the representative of the average Galactic spectra, 
especially for MeV CRs, is still not clear. However, the analysis of gamma 
rays from 
Molecular clouds (e.\,g., \cite{Yang2014}) seems to indicate that at least the 
spectrum of proton CRs of energy above a few GeV is quite homogeneous in our 
Galaxy. The intensity of CRs in the local ISM as revealed by Voyager 
measurements is too weak to explain the level of ionization rate observed in 
clouds. Possible solutions to this problem include the presence of another 
source of ionization or a non-uniform intensity of low energy CRs throughout 
the Galaxy \cite{Phan2018}.

One of the possible sources of CR was suggested by \cite{Padovani2016}. They supposed the existence of an acceleration mechanism for both cosmic-ray protons and electrons through the diffusive shock acceleration near protostars. This mechanism may explain the high ionisation rate and the synchrotron emission observed. Also such internal source of energetic particles should have a strong influence on the ionisation of the protostellar disc, on the star and planet formation processes, and on the formation of pre-biotic molecules \cite{Padovani2016}.

As in the solar system where solar wind can inhibit the CRs propagation, also stellar winds can modulate CRs spectra within the circumstellar environment and subsequently into the disk. In \cite{Cleeves2013} a two-dimensional protoplanetary disk model of a T-Tauri star system was constructed and ionization from stellar and interstellar far ultraviolet, stellar X-ray photons, and CRs was found. According to their results stellar winds can conform a heliosphere-like analog, i.\,e., using the term of the authors “T-Tauriosphere” decreases CR ionization rates by several orders of magnitude at low to moderate CR energies ($\sim 1$~GeV). The implications of a diminished CR ionization rate on the physics of the disk was performed by investigation of the magnetorotational instability: if winds are so efficient at CR modulation, than the major source of ionization would be short-lived radionuclides \cite{Cleeves2013}. However, the authors didn’t consider the stellar CR.

Chandra X-ray satellite observed the large flares on T Tauri stars which can give a significant amount of energetic particles and recently the Herschel Space Observatory provided evidence of their possible contribution to the disk ionization of young stars. Authors of \cite{Rab2017} modeled ionization of stellar energetic particles in protoplanetary disks around T Tauri stars using a particle energy distribution derived from solar flare observations and an enhanced stellar particle flux proposed for T Tauri stars. They have shown that stellar energetic particles can be an important ionization agent for disk chemistry.

Test-particle numerical simulations of energetic protons propagating into a realistic T Tauri stellar wind have shown that in the energy range 0.1--10~GeV they are consistent with expectations from Chandra and the Herschel Space Observatory observations \cite{Fraschetti2018}. Also in \cite{Fraschetti2018} was found that the disk ionization is dominated by X-rays over much of its area, except some narrow regions where particles are channeled by the turbulent magnetic field, in contrast with a previous theoretical study. 

Moreover, a proper description of the particle transport is essential to compute the ionization rate since the electron and positron differential fluxes depend sensitively on the fluxes of both protons and photons \cite{Padovani2018}. Their results show that the CR ionization rate in high-density environments, such as the inner parts of collapsing molecular clouds or the mid-plane of circumstellar disks, is higher than previously assumed. 

The exoplanets discovery gives a great impetus to investigation both the habitability conditions and possible habitable planets. However, from the investigation of the Earth environment it seems to be clear that the radiation environment should have great influence on any planet and its atmosphere. Despite of their importance CRs and their influence on the radiation conditions near planets have dropped out of the discussion, and, perhaps, only one group investigate the influence of CRs on exoplanets (see for example \cite{Griessmeier2005}). The influence of Galactic cosmic rays (GCRs) on a planets depend on planetary magnetosphere and atmosphere and has been considered in several papers (see, e.g., \cite{Atri2013,Griessmeier2015}). In paper \cite{Griessmeier2016} the influence of CRs on the atmosphere near an Earth-like exoplanet was investigated and it was suggested that the GCR flux might be considered as an isotropic and nearly constant one. However the authors did not consider GCR modulation.

Such problem is similar to the investigations of archean Earth in the young Sun environment. For example, quantitative model developed in \cite{Scherer2002} demonstrated that a change of the interstellar medium surrounding the heliosphere triggers significant changes of planetary environments caused by enhanced fluxes of neutral atoms as well as by the increased cosmic ray fluxes for the times 10$^{4}$--10$^{6}$  years. In paper \cite{Cohen2012} the 3D MHD models for solar wind and 2D model for CGR transferring near archean Earth were used. It was showed that the GCR flux would has greatly reduced with comparison to the present conditions. The reason of it is mainly due to the shorter solar rotation period and tighter winding of the Parker spiral, and to the different surface distribution of the more active solar magnetic field.

In contrast to Galactic cosmic rays, the detection (or separation from the 
total flux) of SCR is impossible far from the parent star, and because of 
this they are the ``stepsons'' of CR physics usually mentioned as a possible 
CR component approximately once every ten years \cite{Unsold1957,Edwards1971,Lovell1974,Mullan1979,Kopysov2005,Stozhkov2011,Struminsky2017}. Nevertheless, stellar CRs 
are considered as an important factor of space weather in the habitable zone 
of a star in many papers \cite{Tabataba2016,Atri2017,Struminsky2017}. Unfortunately, the spectrum of stellar CRs 
cannot be determined and the wellknown spectra of solar CRs \cite{Atri2017,Tabataba2016} are used to model stellar CRs. Another approach 
is based on general physical laws and also uses the solar--stellar analogy, 
but without using near-Earth solar CR observations \cite{Struminsky2017}.

Here we review methods proposed in our previous papers \cite{Struminsky2017,SS18PAG,SS18GA} showing the Sun-stellar similarities and scaling of stellar parameters which allows us to obtain the correct estimates for different stars by the order-of-magnitude. The previous paper was devoted to the Proxima Centauri system \cite{SS18PAG} and for TRAPPIST-1 system \cite{SS18GA}. Below we present some our results for TRAPPIST-1 and new unpublished estimates for $\sigma$ Ori E. The next section is devoted to the methods using for calculations and their discussion. The section 3 presents results for TRAPPIST and $\sigma$~Ori~E.

\section{Methods}

\subsection{Stellar wind and astrosphere}

Solar wind velocity, particle density, magnetic field and dimension of the 
heliosphere are essential parameters for propagation and modulation of CR 
observed near the Earth. Analogous parameters of stellar wind and 
astrosphere would be important for CR near exoplanets of some chosen star.

For the stellar wind velocity, we use the Parker's model \cite{Parker1958}, 
which depends only of the coronal temperature and star radius and that's why 
is universal for any star system. For a given coronal temperature $T_{e}$ we 
may estimate the sound speed $u_{cr}=({2k_{B}T_{e}} / m_{p})^{-1/2}$, critical distance, where the wind speed became equal to the sound speed, $r_{cr}={GM_\ast}/u_{cr}$ (here $M_\ast$ is a star mass, $G$ is the 
gravitational constant). The stellar wind speed for the distance much larger 
than critical may be estimated as $V_{\text{SW}}=u_{cr}\log(r/r_{cr})$.

We may get a coronal temperature from X-ray observations of particular star 
or may use for estimates a maximal possible value of the coronal temperature 
$$
T_{\text{cor}}^{\max}=G\frac{M_{\ast }m_{p}}{4kR_{\ast}}(\sqrt 3 -1)
$$
determined by the condition that a radius of critical point is equal to 
the stellar radius plus the coronal height 
$$
H_{\text{cor}}=\frac{2kT_{\text{cor}}m_{p}R_{\ast }^{2}}{GM_{\ast }}.
$$
Here $R_\ast$ is the star radius.

The number density of stellar wind may be estimated from the thermal coronal 
loss. Let us suppose that the heat flux from corona is $Q=-({8\pi 
}/{7})R_{\ast }k(T_{e})T_{e}$, where $k\left( T_{e} \right)=6\times
{10}^{-6}T_{e}^{5/2}$~$\text{erg}\cdot\text{cm}^{-1}\times\text{s}^{-1}\cdot\text{K}^{-1}$ is the thermal 
conductivity coefficient for fully ionized gas. Assuming that all the heat 
flux $Q$ goes for the coronal spherical symmetrical expansion, we find that
$$
Q\approx 4\pi r^{2}m_{p}\frac{N(r)V_{\text{SW}}}{2}(V_{\text{SW}}^{2}+V_{\text{esc}}^{2}),
$$ 
where 
$V_{\text{esc}}=({2GM_{\ast}}/{R_{\ast }})^{1/2}$ is the escape velocity) and 
stellar wind number density $N$ at the distance $r$ is
\[
N(r)=\frac{48\pi }{7}R_{\ast }\frac{{10}^{-6}T_{e}^{7/2}}{2\pi 
	r^{2}m_{p}V_{\text{SW}}\left( V_{\text{SW}}^{2}+V_{\text{esc}}^{2} \right)}
\]
Also now it is possible to find the mass loss rate $M_{r}=4\pi 
r^{2}Nm_{p}V_{\text{SW}}$ and the astrosphere radius 
$$
R_{\text{as}}=R\sqrt 
\frac{m_{p}NV_{\text{SW}}^{2}}{P_{\text{ISM}}}, 
$$
where $P_{\text{ISM}}=1$~eV/cm$^{3}$ is the 
pressure of interstellar medium. 

For estimates of magnetic field in some point of the astrosphere we may use 
the Parker spiral. From the magnetic flux conservation follows $B(r)=B_{\ast 
}\left({R_{\ast }}/{r} \right)^{2}$, where $B_{\ast }$ is the stellar 
magnetic field at photospheric level. The radius of the first turn of Parker 
spiral equals to $r_{1}=V_{\text{SW}}T({\varphi }/{2\pi })$, where $T$ is the 
stellar rotation period.

\subsection{CR modulation}
\label{sec:mylabel1}
According to \cite{Parker19581} a modulated flux of galactic cosmic rays is
\[
j_{0}\left( \eta \right)=j_{\infty }\left( \eta \right)\exp\left\{ 
-\frac{12V_{\text{SW}}\left( r_{1}-r_{2} \right)lZ^{2}e^{2}B^{2}(\eta +1)}{\pi 
	^{2}m^{2}c^{5}\left[ \eta (\eta +2) \right]^{3/2}} \right\},
\]
Where $\eta $ is a ratio of kinetic energy to the rest particle energy, 
$Ze$ is a particle charge, $m$ is the particle mass, $l$ is a scale of magnetic field fluctuations ($l=2\times 
{10}^{6}$~km for the solar wind), $\left( r_{1}-r_{2} \right)$ is a dimension of modulation shell. In our 
estimates we assume that a dimension of modulation shell is a difference 
between radius of inneresting point and a radius of first Parker turn. In 
this case a radius of astrosphere should be greater than a radius of first 
Parker turn and the planet orbit should be within the first turn. Therefore 
to estimate CR modulation near exoplanets we need to know magnetic field and 
rotation period from observations of hosting star. 

\subsection{Stellar cosmic rays}
To estimate the SCR influence let us assume that in stellar flares mechanism 
of protons acceleration is the same as in the solar flares and the main 
buildings blocks in star and solar coronas are magnetic loops \cite{Strumisky2017}.

Our main suggestion is that in the stable loop magnetic field energy should 
be in equipartition with thermal plasma energy. From pressure balance 
$B_{0}^{2} /{8\pi =nkT={Gm_{p}M_{\ast }H} /R_{\ast }^{2}}$ assuming the mean free path $H = (n\sigma 
_{T})^{-1}$ (where $\sigma_{T}$ is the Thompson scattering 
cross-section) we find estimation of photospheric magnetic field in the form 
$$
B_{0}=\frac{1}{R_{\ast }}\sqrt \frac{8\pi Gm_{p}M_{\ast }}{\sigma_{T}}. 
$$ 

Balona \cite{Balona2015} showed that the characteristic scale of an active region is 
about tenth percent of star radius, $L=\alpha R$, magnetic field over an 
active region is about one tenth of photospheric $B=\beta B_{0}$. Here 
coefficients $\alpha \beta <1$. Using these parameters we can estimate the 
flare energy as 
$$
E_{fl}=\frac{B^{2}L^{3}}{8\pi }=2.3\times {10}^{37}\alpha^{3}\beta^{2}\left( \frac{R}{R_{\odot}} \right)\left( 
\frac{M}{M_{\odot}} \right).
$$ 
In big solar flares approximately 10{\%} of flares 
energy is the energy of protons acceleration. Assuming similar processes for 
the stars we estimate amount of accelerated protons with average energy 
$E_{p}$ as $N=0.1E_{fl}/E_{p}$.

Electric field over an active region is $E=uB/c$, where $u$ is 
velocity of reconnection, and the maximum energy of protons accelerated in 
flares would be $E_{\max}=({\alpha \beta }/{c})uB_{0}R_{\ast }$. 

The largest fluxes of solar protons $j_{\max}$ observed near the Earth 
correspond to arrival of strong magnetic field disturbances and 
$j_{\max}=F_{\max}/(2\pi \tau)$, where $\tau =r/V_{\text{SW}}$ is a propagation 
time of solar wind disturbances to 1~AU. Let us assume a similar values for 
stellar cosmic rays, i.\:e., assume that the convection plays a main role in propagation of 
stellar CR for strong magnetic field. Note that the solar magnetic field is 
rather weak in comparison with magnetic field of several tens to several 
thousand Gauss observed at magnetic active stars. Considering that stellar CR 
propagate within spatial angle of $60\times60$ degrees we will get fluencies 
$F=9N/\pi r^{2}$ and maximal intensities $j_{\max}=F/(2\pi \tau)$, which might 
be observed at distance $r$ in the astrosphere.

\subsection{Discussion of methods}
According to the dynamo theory O--B, A stars do not have developed convective zone 
	contrary to F--M stars and, therefore, they should not have magnetic field. 
	If there are no magnetic fields, then would be absent stellar activity, 
	flares, hot corona and hot stellar wind. However magnetic fields 
	\cite{Berdyugina2009,Linsky2015}, X-ray emission (see \cite{Gudel2004} and 
	references therein) and flares \cite{Groote2004} are observed at O--B 
	stars, flares \cite{Balona2012} and stellar spots \cite{Balona2017} are observed 
	at A-stars. A nature of magnetic fields is still unknown \cite{Hubrig2015}, X-ray emission is described by shock waves propagating in cold 
	stellar wind \cite{Babel1997} and stellar flares are attributed to 
	invisible cold companion \cite{Groote2004,Mullan2009,Pedersen2017} or wind-shock X-ray emission \cite{Groote2004}. If 
	independently on magnetic field nature for rather strong magnetic fields 
	suppose that magnetic structures like arks and filaments could be formed at O--B, 
	A stars, then we may discuss expanding hot coronas (stellar wind) and their 
	X-ray emission, stellar flares, do estimates using solar-stellar analogous 
	\cite{SS18PAG}.

\section{Results}

\subsection{Estimates for TRAPPIST 1}

In the year of 2016 the planetary system of TRAPPIST-1 with 7 
exoplanets was discovered, four of them are in a possible habitable zone \cite{Gillon2017}. TRAPPIST-1 is a M8 star at distance of 10 light years, its parameters 
are $M =(0.089\pm 0.006) M_{\odot }$, radius $R = (0.121 \pm 0.003)R_{\odot }$, effective temperature $T = 2516$~K. The rotation period is 
$P_{\text{rot}}  = 3.295\pm 0.003$~days \cite{Vida2017}. The \mbox{TRAPPIST-1} star 
has a hot corona with ratio of X-ray to bolometric luminosity 
$L_{x}/L_{\text{bol}}  = (2$--$4) \times 10^{-4}$ \cite{Wheatley2017}, 
i.e., should have a hot stellar wind similar to the solar wind. Its flare 
activity was observed by the MOST satellite \cite{Vida2017} a frequency 
of flares with energy $\sim10^{33}$~erg is $f= 1.2\times 
10^{-7}$~s$^{-1}$. The measured average magnetic field was estimated as 600~G \cite{Reiners2010}, but it is a too weak source to get ZDI magnetic 
field maps by using modern instruments. In this paper we will not discuss 
possibility of generation of such great field, which are really doubly.

In our recent paper \cite{SS18GA} we presented our view on radiation conditions in the TRAPPIST 1 system. Here we show some our estimates only for TRAPPIST 1d, one of four exoplanets in the habitable zone. 

The modeling characteristic values for stellar wind of TRAPPIST 1 are shown 
in Tab. 1 for different values of the coronal temperature.

	\begin{table}[htbp]
	\caption{Stellar wind parameters for the TRAPPIST-1 d ($r_{d}=0.0214$~AU)}
\begin{center}
	\def\arraystretch{1.4}
		\begin{tabular}{|l|l|l|l|l|l|l|l|}
			\hline
			$T$, MK& 
			$Q$, erg/s& 
			$n_{\text{SW}}$, cm$^{-3}$& 
			$u_{cr}$, km/s& 
			$r_{cr}$/R$_{\ast }$& 
			$V_{\text{SW}}$, km/s& 
			$r_{\text{Park}}$, AU& 
			$R_{\text{AS}}$, AU \\
			\hline
			1& 
			$1.813\times10^{26}$& 
			1614& 
			131& 
			4.0& 
			294/366& 
			0.6& 
			61.5 \\
			\hline
			2& 
			$2.051\times10^{27}$& 
			6267& 
			186& 
			2.0& 
			545/647& 
			1& 
			224 \\
			\hline
			3& 
			$8.480\times10^{27}$& 
			12510& 
			228& 
			1.4& 
			760/884& 
			1.5& 
			442 \\
			\hline
			4& 
			$2.321\times10^{28}$& 
			19710& 
			263& 
			1.0& 
			953/1097& 
			1.8& 
			696 \\
			\hline
		\end{tabular}
\end{center}
		\label{tab1}
	\end{table}

Figure 1 illustrates a possible effect of GCR modulation for hypothetical 
values of TRAPPIST 1d magnetic field of 1--600~G and coronal temperatures of 
2~MK. This value of coronal temperature corresponds to the solar coronal 
temperature and is about twice less than the maximum possible temperature 
for TRAPPIST 1 corona. From Fig.~1 it is clear, that CR with energies less 1~TeV should be swept out from the astrosphere in a case of magnetic field 
$>300$~G. Moreover as $R_{AS}\gg r_{1}$ the real modulation GCR will 
be greater. However it is possible, that the real magnetic fields will be 
much less than the measured. The modulation effects even for 1~G are more 
than the order of magnitude higher than for the Earth (Fig.~1). 
\begin{figure}
	\begin{center}
		\includegraphics{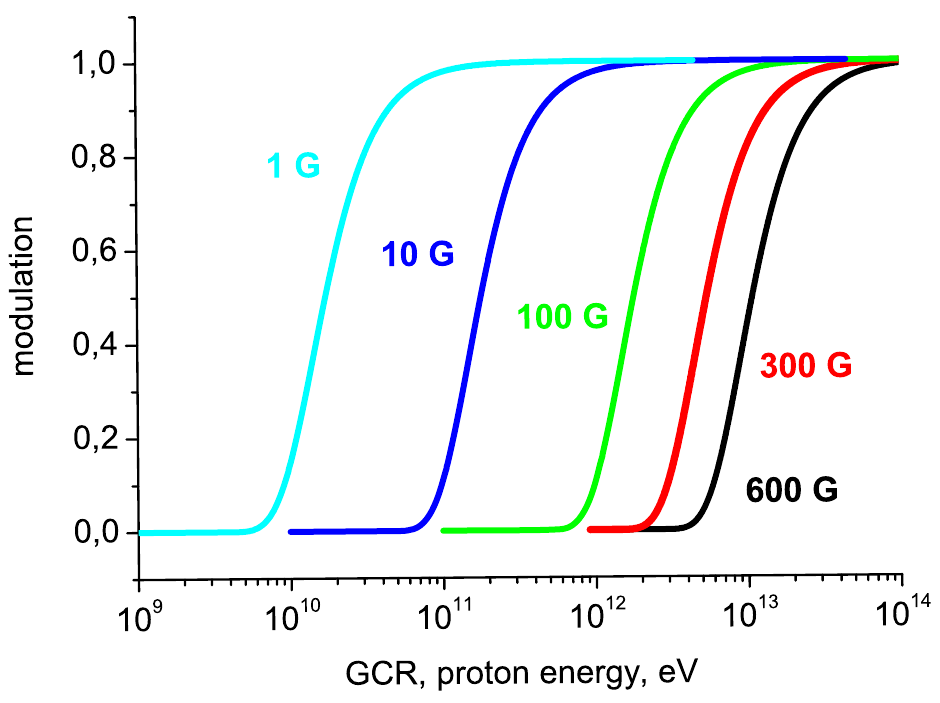}
	\end{center}
	\caption{\label{fig1}The dependence of CR modulation for TRAPPIST 1d for different values 
		of stellar magnetic field and stellar wind velocity 545~km/s ($T_{\text{cor}}=2$~MK)}
\end{figure}

Assume the following parameters of active region: $L=\alpha R=\alpha 
8.5\times10^{9}$~cm, $B = \beta B_{0}=3000\beta$~G, $V = 100$~km/s, we find flare 
energy $E_{fl}\alpha^{3}\beta^{2}= 2.2\times10^{35}\alpha^{3}\beta^{2}~\text{erg}=10^{33}$~erg which is a reasonable value for 
TRAPPIST 1 flares. A frequency of such flares is $f = 10^{-2}~\text{year}^{-1}=1.2^{-7}$~s$^{-1}$. The number of protons accelerated in one flare and average 
rate of their generation as well as their density and flux within the first 
Parker spiral turn are presented in Tab.~2.

	\begin{table}[htbp]
\caption{The number of protons $N$ accelerated in one flare and average rate of 
	their generation, density and flux of protons within the first Parker spiral 
	turn (1~AU)}
\begin{center}
	\def\arraystretch{1.4}
		\begin{tabular}{|l|l|l|p{80pt}|p{90pt}|}
			\hline
			$E_{p}$, MeV& 
			$N$, protons& 
			$fN$, proton/s& 
			$n={3N}/{4\pi r_{\text{park}}^{3}}$, proton/cm$^{3}$& 
			$nV_{\text{SW}}/(2\pi)$, proton/(cm$^{2}\cdot\text{s}\cdot\text{sr}$) \\
			\hline
			30& 
			$2.1\times 10^{36}$& 
			$2.5\times 10^{29}$& 
			$1.5 \times 10^{-4}$& 
			1300 \\
			\hline
			200& 
			$3.1\times 10^{35}$& 
			$3.7\times 10^{28}$& 
			$2.2 \times 10^{-5}$& 
			191 \\
			\hline
		\end{tabular}
		\label{tab2}
	\end{center}
\end{table}

Maximum proton fluences and intensities for possible extreme events of stellar CR, Trappist-1d are presented in Tab.~3. 

\begin{table}[htbp]
\caption{Maximum fluences and proton intensity for $V_{\text{SW}}=545$~km/s in 
	extreme events near TRAPPIST 1 d ($r_d = 0.0214$~AU)}
\begin{center}
	\def\arraystretch{1.4}
		\begin{tabular}{|l|l|l|l|}
			\hline
			$E_{p}$ , MeV& 
			$N$, protons& 
			$F=9N/\pi r^{2}$, cm$^{-2}$& 
			$j_{\max}=F/(2\pi \tau )$ ($\text{cm}\cdot\text{s}\cdot\text{sr})^{-1}$ \\
			\hline
			30 & 
			$2.1\times 10^{36}$& 
			$5.8\times 10^{13}$& 
			$1.6\times 10^{9}$ \\
			\hline
			200 & 
			$3.1\times 10^{35}$& 
			$8.6\times 10^{12}$& 
			$2.4\times 10^{8}$ \\
			\hline
		\end{tabular}
		\label{tab3}
	\end{center}
\end{table}

\subsection{Estimates for $\sigma $ Ori E}
The magnetic helium-strong star $\sigma $ Ori E is famous for its 
	outstanding characteristics: the surface magnetic field strengths are at least 10~kG 
	\cite{Landstreet1978}, X-ray emission $L_{X}/L_{b}=10^{-7}$ 
	\cite{Pallavicini1981} and $L_{X}/L_{b}=3.9\times 10^{-7}$, 
	flare activity \cite{Groote2004,Mullan2009}, a weak cold stellar 
	wind of some $10^{-10}M_{\odot}$~year$^{-1}$ \cite{Groote1982}.
Fundamental parameters of $\sigma$ Ori E are $T_\text{eff} = 22500$~K, $R=3.77R_{\odot}$, $M=8.3M_{\odot}$, the rotation period $P=1.190847$~days \cite{Townsend2013}. The estimated overall energy of X-ray flare observed at 
$\sigma $ Ori E was between $5.3\times 10^{35}$ and $2.9\times 10^{36}$ \cite{Groote2004}.

Applying the methods presented above we may get values of X-ray luminosity  
$L_{x} $ and flare energy $\alpha^{3}\beta^{2}E_{fl}$, which 
are very close to the observed. The mass rate of hot stellar wind 
appeared to be about 3--4 orders less than the observed mass rate of cold 
stellar wind $10^{-10} M_{\odot}$~year$^{-1}$ \cite{Groote1982}, 
that may explain why the hot stellar wind of $\sigma$ Ori~E is not 
observed. An electron density in the corona $n_{e}$ was a free parameter 
in our calculations.

\begin{table}[htbp]
\caption{Calculated parameters of the hot corona of $\sigma 
$ Ori E}
		\begin{center}
			\def\arraystretch{1.4}
			\begin{tabular}{|l|l|l|l|l|l|p{66pt}|}
			\hline
			$T$, MK& 
			$n_{e}$, cm$^{-3}$& 
			$H_{\text{cor}}/R_{\ast}$& 
			$L_{X}/L_{b}$& 
			$M_{r}$ ($M_{\odot}$, year$^{-1}$)& 
			B$_{0}$, G& 
			$\alpha^{3}\beta^{2}E_{fl}$, erg \\
			\hline
			4.6& 
			$2.9\times 10^{8}$& 
			0.18& 
			$1.4\times 10^{-7}$& 
			$4.0\times 10^{-14}$& 
			980& 
			$4.9\times 10^{38}$ \\
			\hline
			8.9& 
			$2.6\times 10^{8}$& 
			0.37& 
			$1.3\times 10^{-7}$& 
			$1.7\times 10^{-13}$& 
			980& 
			$6.9\times 10^{38}$ \\
			\hline
		\end{tabular}
		\label{tab4}
	\end{center}
\end{table}

\begin{table}[htbp]
	\caption{Stellar wind parameters for $\sigma $ Ori E in the 
		habitable zone ($\sim14$~AU)}
\begin{center}
	\def\arraystretch{1.4}
		\begin{tabular}{|l|l|l|l|l|l|l|l|}
			\hline
			$T$, MK& 
			$Q$, erg/s& 
			$n_{\text{SW}}$, cm$^{-3}$& 
			$u_{cr}$, km/s& 
			$r_{cr}/R_{\ast }$& 
			$V_{\text{SW}}$, km/s& 
			$r_{\text{park}}$, AU& 
			$R_{\text{AS}}$ AU \\
			\hline
			4.6& 
			$1.2\times 10^{30}$& 
			0.5& 
			218& 
			2.6& 
			1609& 
			1.1& 
			4051 \\
			\hline
			8.9& 
			$1.2\times 10^{31}$& 
			1.6& 
			392& 
			1.4& 
			2497& 
			1.7& 
			10770 \\
			\hline
		\end{tabular}
		\label{tab5}
	\end{center}
\end{table}

\begin{figure}[htbp]
\begin{center}
		\includegraphics{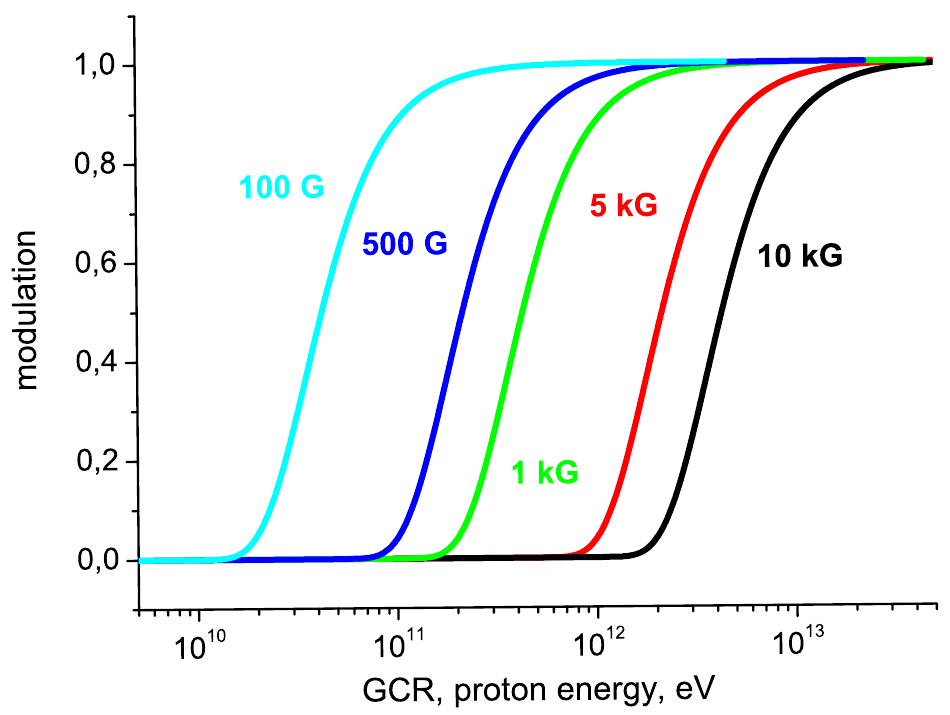}
	\caption{\label{fig2}The dependence of CR modulation in habitable zone of 
		$\sigma $ Ori E for hypothetical values of magnetic field of 100~G--10~kG and 
		coronal temperatures of 4.6~MK}
\end{center}
\end{figure}

Figure 2 illustrates a possible effect of GCR modulation for hypothetical 
values of $\sigma $ Ori E magnetic field of 100~G--10~kG and 
coronal temperatures of 4.6~MK. This value of coronal temperature is about 
twice less than the maximum possible temperature for $\sigma
$ Ori E corona. Since the radius of habitable zone $R_{H}\approx14~\text{AU}\gg r_{\text{park}}\sim1$~AU we assumed a dimension of modulation region is about 30~AU. From Fig.~2 it 
is clear, that CR with energies less 1~TeV should be swept out from the 
astrosphere in a case of magnetic field $>5$~kG. Moreover as 
$R_{\text{AS}}\gg30$~AU the real modulation GCR will be greater. However it is possible, that 
real values of the stellar magnetic fields will be much less than the 
measured. Flare observations may help to estimate their real values.

Assuming the following parameters of active region: 
$L=H_{\text{cor}}=\alpha R=0.18R(0.37R)$, $B = \beta B_{0}=980\beta$~G, $V = 100$ km/s, we 
find flare energy $E_{fl}\alpha^{3}\beta^{2}=2.8\times10^{36}\beta^{2}$ ($3.4\times 10^{37}\beta^{2}$)~erg, 
which is very close to the observed flare energy $5.3\times 10^{35}$--$2.9\times 10^{36}$~erg \cite{Groote2004}, i.e., $\beta=0.44$--0.29. The number of protons accelerated in such a flare as well as their 
density and flux within the radius of habitable zone are presented in Tab. 
6. These values are in agreement with our previous estimates and conclusions 
\cite{Struminsky2017}.

\begin{table}[htbp]
	\caption{Number of protons, their maximum fluencies and intensities in 
		habitable zone of $\sigma$Ori E ($R_H=14$~AU, flare 
		energy $2\times 10^{36}$~erg)}
\begin{center}
	\def\arraystretch{1.4}
			\begin{tabular}{|l|l|l|l|}
			\hline
			$E_{p}$ , MeV& 
			$N$, protons& 
			$F=9N/\pi r^{2}$, cm$^{-2}$& 
			$j_{\max}=F/(2\pi \tau )$, $\text{cm}\cdot\text{s}\cdot\text{sr}^{-1}$ \\
			\hline
			30 & 
			$4.2\times 10^{40}$& 
			$2.7\times 10^{12}$& 
			$3.2\times 10^{5}$ \\
			\hline
			200 & 
			$6.2\times 10^{39}$& 
			$4.1\times 10^{11}$& 
			$4.9\times 10^{4}$ \\
			\hline
		\end{tabular}
		\label{tab6}
	\end{center}
\end{table}

It is possible that the observed magnetic field at $\sigma 
$  Ori E $B_{\ast }=10$~kG might be not the average stellar 
magnetic field but the local one (magnetic field of active region). In this 
case we may right $\beta_{\ast }B_{\ast }=\beta B_{0}$ and $\beta 
_{\ast }=\beta B_{0}/B_{\ast }=0.04$--0.03. An average magnetic field should 
be much less than 10~kG, about several tens of Gauss, that is below the 
observational threshold. 

\section{Conclusions}

It may be said recent measurements by the Voyager I spacecraft give the birth of qualitative low energy CR astrophysics, have provided data of lower-energy cosmic-ray particles out of heliosphere \cite{Stone2013}. This give us the reference point to evaluate low-energy CR intensity obtained by other methods in molecular clouds, protoplanetary discs and astrospheres. Our knowledge of the low CR interaction environment is at the same level as our knowledge of the heliosphere in the 1950s. So for the first and simple estimates we can use the Sun-stellar similarities and the methods developed for the CR interaction with heliospheric medium. The most unknown parameter in these estimates is stellar magnetic field which can be different by several orders of magnitude in comparison with solar magnetic field. The main problem is the measured stellar field may be magnetic field of active regions, i.e., the local one. We believe that developed approach can be used for any magnetic star with hot corona. Results for and any planetary system should coincide with those obtained by more advanced models by the order of magnitude. 

The estimations for stellar wind parameters, GCR and SCR parameters were found for TRAPPIST 1 system. We obtain that for corona temperature 2~MK the scale of astrosphere is $\sim224$~AU and radius of the first Parker spiral is $\sim1$~AU. Particles should be swept out by the stellar wind and the main factor of GCR modulation is the stellar wind magnetic field. As a result of modulation GCR less than 1 TeV should be absent near planet. However, flares frequency and their energy allow us to state that the radiation environment determined by SCR.

For $\sigma$ Ori E applying the methods presented above we obtain the values of X-ray luminosity $L_X$ and flare energy $\alpha^3\beta^2E_{fl}$, which are very close to the observed. The mass rate of hot stellar wind appeared to be about 3--4 orders less than the observed mass rate of cold stellar wind $10^{-10}M_\odot$~year$^{-1}$ \cite{Groote1982}, that may explain why the hot stellar wind of $\sigma$ Ori E is not observed. From the flares’ energy estimations it is possible that the observed magnetic field at $\sigma$ Ori E might be not the average stellar magnetic field but the local one. 

e-ASTROGAM is a concept of a breakthrough observatory carrying a $\gamma$-ray telescope for the study of the non-thermal Universe in the photon energy range from 0.15 MeV to 3 GeV. The gamma-rays of such energy give us the new source for understanding of low-energy CR.

The gamma-rays of energies from 0.15 MeV to 3 GeV may give us information for better understanding of low-energy CR in their different locations in the Galaxy. e-ASTROGAM is a concept of a breakthrough observatory carrying a $\gamma$-ray telescope for the study of the non-thermal Universe in this new photon energy range \cite{AG,Orlando2018}.

\ack{The work is supported by Russian Foundation for Basic Research grant 16-02-00328.}

\end{document}